\documentclass[%
 reprint,
 amsmath,amssymb,
 aps,
]{revtex4-1}

\usepackage{graphicx}
\usepackage{dcolumn}
\usepackage{bm}

\begin{document}


\title{Non-equilibrium correlations in minimal dynamical models of polymer copying }
\author{Jenny M Poulton}
\affiliation{Department of Bioengineering, Imperial College London, London SW7 2AZ, United Kingdom}%

\author{Pieter Rein ten Wolde}
\affiliation{FOM Institute AMOLF, Science Park 104, 1098 XE Amsterdam, The Netherlands}%

\author{Thomas E Ouldridge}
\affiliation{Department of Bioengineering, Imperial College London, London SW7 2AZ, United Kingdom}%

\date{\today}

\begin{abstract}
Living systems produce ``persistent" copies of information-carrying polymers, in which template and copy sequences remain correlated after physically decoupling. 
We identify a general measure of the thermodynamic efficiency with which these non-equilibrium states are created, and analyze the accuracy and efficiency of a family of dynamical models that produce persistent copies. For the weakest chemical driving, when polymer growth occurs in equilibrium, both the copy accuracy and, more surprisingly, the efficiency vanish. At higher driving strengths, accuracy and efficiency both increase, with efficiency showing one or more peaks at moderate driving.  Correlations generated within the copy sequence, as well as between template and copy, store additional free energy in the copied polymer and limit the single-site accuracy for a given chemical work input. Our results provide insight in the design of natural self-replicating systems and can aid the design of synthetic replicators.
\end{abstract}

\pacs{Valid PACS appear here}
\maketitle

\begin{figure}
\includegraphics[scale=0.75]{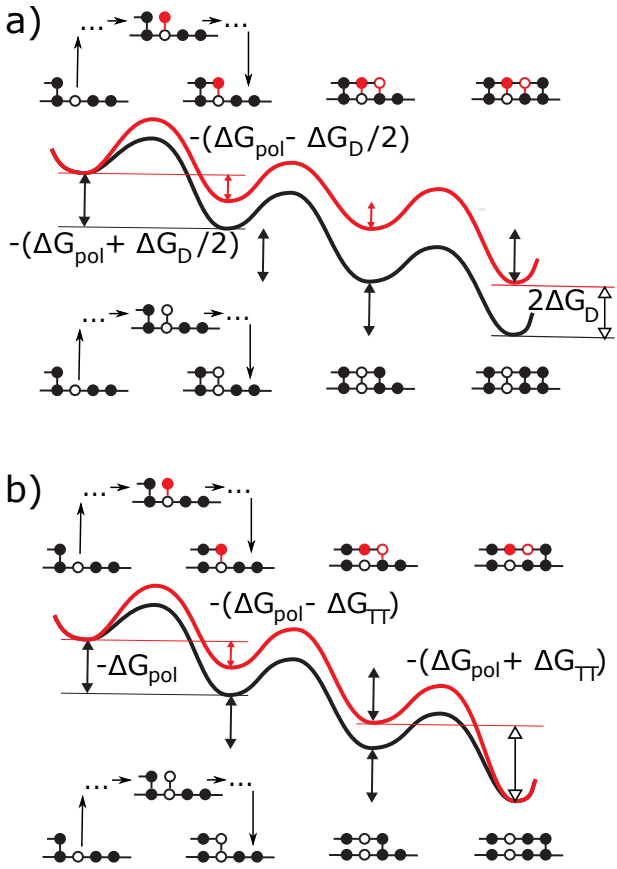}
\caption{Free-energy landscapes for simple examples of (a) templated self-assembly, in which the monomers remain bound to the template during the copy process; and (b) persistent copying, in which the monomers detach from the template after they have been incorporated into the polymer. Both diagrams show the addition of three monomers to a growing polymer, driven by a chemical free energy of backbone polymerisation $\Delta G_{\rm pol}$. In each case, two scenarios are considered: the addition of two incorrect monomers, followed by a correct one (top), and the addition of three correct monomers (bottom). Local minima in the landscape represent macrostates following complete incorporation of monomers; intermediate configurations, illustrated schematically for the first transition, are part of the effective barriers. 
In templated self-assembly,
the chemical free-energy cost of previously incorporated mismatches is retained as the daughter grows \cite{Bennett,Cady,Andrieux,Sartori1,Sartori2,esposito2010}. Thus in (a), each mismatch in the daughter increases the chemical free energy by $\Delta G_{\rm D}$ relative to the perfect match. 
In persistent copying (b), the chemical free-energy penalty for incorporating wrong momomers is only temporary; it arises when the incorrect monomer is added to the growing polymer, but is lost when that monomer subsequently detaches from the template. As a result, the overall chemical free-energy change of creating an incorrect polymer is the same as that for a correct one. Analyzing the consequences of this constraint, which is a generic feature of copying but does not arise in TSA, is the essence of this work. The figure also shows that in our specific model, incorporating a wrong monomer after a correct one tends to reduce the chemical free-energy drop to \(\Delta G_{\rm pol}-\Delta G_{\rm TT}\), and incorporating a correct monomer after an incorrect one tends to increase it to \(\Delta G_{\rm pol}+\Delta G_{\rm TT}\); however, adding a wrong monomer to a wrong one, and adding a correct monomer to a correct one, does not change the free-energy drop \(\Delta  G_{\rm pol}\).}

\label{Phase}
\end{figure}

The copying of information from a template into a substrate is  fundamental to life. The most powerful copying mechanisms are persistent, autonomous and generic. A persistent copy retains the copied data after physically decoupling from its template \cite{Ouldridge,OuldridgePRX}. An autonomous copy process does not require systematically time-varying external conditions \cite{OuldridgePRX}, making it more versatile. Finally, a generic copy process is able to copy arbitrary data. DNA replication and both steps of gene expression necessarily exhibit all three characteristics.

Unlike natural systems, synthetic copying mechanisms developed hitherto have not incorporated all three features. Early work focused on using template polymers to synthesize specific daughter polymers, but failed to adequately demonstrate subsequent separation of copy and template \cite{Tjivikua,Feng}. 
We describe such a process as templated self-assembly (TSA), by analogy with structures that assemble with high specificity due to favourable contacts in the final state.

Due to cooperativity, the tendency of copies to remain bound to templates grows with template length \cite{Vidonne,orgel1992,colomb2015}. Consequently, generic copying of long polymers (as opposed to dimers and trimers \cite{Sievers,Vidonne,Lincoln}) has proved challenging. One tactic is to consider environments in which the system experiences cyclically varying conditions, with assembly of the copy favoured in one set of conditions and detachment from the template in another
\cite{Wu,Wang,Li}.
A more subtle approach is to use a spatially non-uniform environment, so that individual molecules undergo cyclic variation in conditions \cite{Mast2010}.
Whilst these experiments may indeed reflect early life \cite{Orgel,martin}, they do not demonstrate copying in a truly autonomous context. 

We also contrast the copying of a generic polymer sequence with the approach in Refs. \cite{Schulman,colomb2015}.
Here, the information is propagated between successive units of a single self-assembling polymer, rather than between a template and a daughter polymer, limiting information transmission. Externally induced mechanical stress on long length scales severs the polymers, leading to more nucleation sites and exponential growth.

These challenges suggest that a full understanding of the basic biophysics of copying is lacking. Recently, we outlined fundamental thermodynamic constraints imposed by persistence \cite{Ouldridge}, but did not propose a dynamical mechanism for autonomous copying. Previous analyses fall into two major categories: those that remain agnostic about the distinction between TSA and copying by considering thermodynamically self-consistent models for only part of the polymerization process \cite{Hopfield}, and those that explicitly
address TSA \cite{Bennett,Cady,Andrieux,Sartori1,Sartori2,esposito2010,EHRENBERG1980333,Johansson}.

In this work we analyze a family of model systems that generate persistent copies in an autonomous and generic way. We introduce a new metric for the thermodynamic efficiency of copying, and investigate the accuracy and efficiency of our models. We highlight the profound consequences of requiring persistence, namely that correlations between copy and template can only be generated by pushing the system out of equilibrium. Previous work has considered self-assembly \cite{Whitelam,nguyen1,Gaspard} or templated self-assembly \cite{Bennett,Cady,Andrieux,Sartori1,Sartori2,Gaspard,esposito2010,EHRENBERG1980333,Johansson} in non-equilibrium contexts; in these cases, however, the non-equilibrium driving merely modulates a non-zero equilibrium specificity. Alongside the effect on copy-template interactions, we find that intra-copy-sequence correlations arise naturally. These correlations store additional free energy in the copied polymer, which do not contribute towards the accuracy of copying.

\section*{Models and Methods}
\subsection*{Model definition}
We consider a copy polymer \({M}=M_{1},...,M_{l}\), made up of a series of sub-units or monomers \(M_{x}\), growing with respect to a template \({N}=N_{1},...,N_{L}\) (\(l\leq L\)). Inspired by transcription and translation, we consider a copy that detaches from the template as it grows; fig. \ref{Phase}b shows the simplest model of this type. We consider whole steps in which a single monomer is added or removed, encompassing many individual chemical sub-steps \cite{Sartori2,Cady}.
After each step there is only a single inter-polymer bond at position \(l\), between $M_l$ and $N_l$. As a new monomer joins the copy at position \(l+1\), the bond position \(l\) is broken, contrasting with previous models of TSA \cite{Bennett,Cady,Andrieux,Sartori1,Sartori2,esposito2010} (Fig.~\ref{Phase}a). Importantly, as explained below, each step now depends on both of the two leading monomers, generating extra correlations within the copy sequence.

\begin{figure}[t]
\includegraphics[scale=0.25]{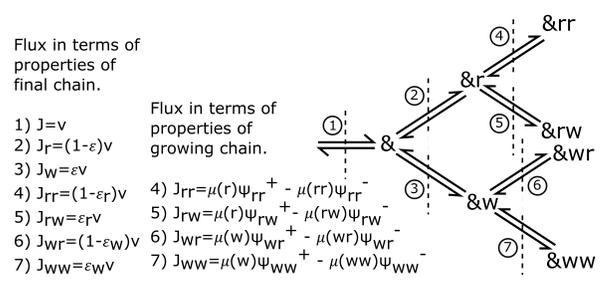}
\caption{Transitions of an arbitrary polymer \(\&\). To relate the final chain to the growing chain it is useful to consider fluxes through interfaces in this transition diagram. Using the tip and combined probabilities, along with relative propensities it is possible to describe the fluxes through interfaces 4-7 in terms of properties of the growing chain. Equally by considering errors and conditional errors and taking fractions of the overall growth velocity, it is possible to find the fluxes through interfaces 4-7 in terms of properties of the final chain and growth velocity.}
\label{Flux}
\end{figure}


Following earlier work, we assume that both polymers are copolymers, and that the two monomer types are symmetric \cite{Bennett,Cady,Andrieux,Sartori1,Sartori2,esposito2010}. Thus the relevant question is whether monomers \(M_{l}\) and \(N_{l}\) match; we therefore ignore the specific sequence of \(N\) and describe \(M_{l}\) simply as right or wrong. Thus \(M_{l}\in r,w\); with example chain \(M=rrwwrrrrrwrr\). An excess of \(r\) indicates a correlation between template and copy sequences. Breaking this symmetry would favour specific template sequences over others, disfavouring the accurate copying of other templates and compromising the generality of the process.

Given the model's state space, we now consider state free energies (which must be time-invariant for autonomy). We treat the environment as a bath of monomers at constant chemical potential \cite{Bennett,Cady,
Andrieux,Sartori1,Sartori2,esposito2010}.
%
%
By symmetry, extending the polymer while leaving the copy-template interaction unchanged involves a fixed polymerization free energy. We thus define \(-\Delta G_{\rm{pol}}\) as the chemical free-energy change for the transition between any specific sequence \(m_{1},...,m_{l}\) and any specific sequence \(m_{1},...,m_{l+1}\), ignoring any contribution from interactions with the template. We then define \(\Delta G_{\rm{TT}}\) as the effect of the free-energy difference between \(r\) and \(w\) interactions with template. This bias can be describes as "temporary thermodynamic" (TT) since it only lasts until that contact is broken.

Overall, each forward  step makes and breaks one copy-template bond. There are four possibilities:  either adding \(r\) or \(w\) at position \(l+1\) to a template with \(M_l = r\); or adding \(r\) or \(w\) in position \(l+1\) to a template with \(M_l = w\). The first and last of these options make and break the same kind of template bond, so the total free-energy change is  \(-\Delta G_{\rm{pol}}\). For the second case there is a \(r\) bond broken and a \(w\) bond added, implying a  free-energy change of  \(-\Delta G_{\rm{pol}}+\Delta G_{\rm{TT}}\).  Conversely, for the third case, there is a \(w\) bond broken and a \(r\) bond added, giving a free-energy change of \(-\Delta G_{\rm{pol}}- \Delta G_{\rm{TT}}\). These constraints are shown in fig. \ref{Phase}b; the contribution of this work is to study the consequences of these constraints. Models of TSA (fig. \ref{Phase}a) of equivalent complexity can be constructed, but they are not bound by these constraints and hence the underlying results and biophysical interpretation are distinct.

Having specified model thermodynamics, we now parameterize kinetics. We assume that there are no ``futile cycles'', such as appear in kinetic proofreading \cite{Hopfield}. 
Reactions are thus tightly coupled: each step requires a well defined input of free energy determined by \(-\Delta G_{\rm{pol}}\) and \(\pm\Delta G_{\rm{TT}}\)  \cite{esposito2018}, and no free-energy release occurs without a step. 

A full kinetic treatment would be a continuous time Markov process  incorporating the intermediate states shown schematically in fig. \ref{Phase}b. To identify sequence output, however, we need only consider the state space in fig. \ref{Phase}b and the relative probabilities for transitions between these explicitly modelled states, ignoring the complexity of non-exponential transition waiting times \cite{Cady}. We define propensities \(\psi^{+}_{xy}\) as the rate per unit time in which a system in state \(\&x\) starts the process of becoming \(\&xy\) and \(\psi^{-}_{xy}\) as the equivalent quantity in the reverse direction (\(\&\) is an unspecified polymer sequence). Our system has eight of these propensities (\(\psi^{\pm}_{rr}\), \(\psi^{\pm}_{rw}\), \(\psi^{\pm}_{wr}\) and \(\psi^{\pm}_{ww}\)); the simplest TSA models require four \cite{Bennett,Sartori1,Sartori2,esposito2010,Cady}.

Prior literature on TSA \cite{Sartori1} has differentiated between purely ``kinetic'' discrimination, in which $r$ and $w$ have an equal template-binding free energy but different binding rates; and 
purely thermodynamic discrimination in which $r$ and $w$ bind at the same rate, but $r$ is stabilized in equilibrium by stronger binding interactions. Eventually, all discrimination is ``kinetic'' for persistent copying, since there is no lasting equilibrium bias (Fig.~\ref{Phase}\,(b)). However, by analogy with TSA, we do consider two distinct mechanisms for discrimination -  a kinetic one, in which \(r\) is added faster than \(w\) to the growing tip, and one based on the temporary thermodynamic bias towards correct matches at the tip of the growing polymer due to short-lived favourable interactions with the template, quantified by \(\Delta  G_{\rm{TT}}>0\) (Fig.~\ref{Phase}b). The kinetic mechanism should not be conflated with fuel consuming ``kinetic proofreading'' cycles that can also enhance accuracy, which are not considered.

We parameterize the propensities as follows. Assuming, for simplicity, that the propensity for adding \(r\) or \(w\) is independent of the previous monomer, we have: \(\psi^{+}_{rr}=\psi^{+}_{wr}\) and \(\psi^{+}_{rw}=\psi^{+}_{ww}=1\), also defining the overall timescale. ``Kinetic'' discrimination is then quantified by
\(\psi^{+}_{xr} /\psi^{+}_{xw}=\exp(\Delta G_{\rm{K}}/k_BT)\). Forwards propensities are thus differentiated solely by \(\Delta G_{\rm{K}}\); backwards propensities are set by fixing the ratios  \(\psi^{+}_{xy}/ \psi^{-}_{xy}\) according to the free energy change of the reaction, which follow from \(\Delta G_{\rm pol}\) and \(\Delta G_{\rm TT}\) (Fig.~\ref{Phase}(b)). Thus, setting \(k_{B}T=1\),
\begin{align}
\begin{split}
  &\psi^{+}_{rr}=e^{\Delta G_{\rm{K}}},\,     \psi^{-}_{rr}=e^{-\Delta G_{\rm{\rm{pol}}}}e^{\Delta G_{K}},\label{eq:rr}
\end{split}\\
\begin{split}
  &\psi^{+}_{rw}=1,\,     \psi^{-}_{rw}=e^{-\Delta G_{\rm{\rm{pol}}}}e^{\Delta G_{\rm{TT}}},\label{eq:rw}
\end{split}\\
\begin{split}
&\psi^{+}_{wr}=e^{\Delta G_{\rm{K}}},\,       \psi^{-}_{wr}=e^{-\Delta G_{\rm{\rm{pol}}}}e^{\Delta G_{\rm{K}}}e^{-\Delta G_{\rm{TT}}},\label{eq:wr}
\end{split}\\
\begin{split}
   &\psi^{+}_{ww}=1,\,        \psi^{-}_{ww}=e^{-\Delta G_{\rm{\rm{pol}}}}.\label{eq:ww}
\end{split}
\end{align}
For a given \(\Delta G_{\rm{K}}\), \(\Delta G_{\rm{\rm{pol}}}\) and \(\Delta G_{\rm{\rm{TT}}}\), eqs. \ref{eq:rr}-\ref{eq:ww} describe a set of models with distinct intermediate states that yield the same copy sequence distribution. We can thus analyze the simplest model in each set, which is Markovian at the level of the explicitly modelled states with \(\psi^{\pm}_{xy}\) as rate constants. 

\subsection*{Model analysis} 
We use Gaspard's method to solve the system \cite{Gaspard}; the underlying equations can also be mapped to models of distinct physical systems that have different constraints on the parameters \cite{Whitelam}. In this approach, the tip-monomer identity probabilities \(\left(\mu (m_{l})\right)\), the joint tip and penultimate monomer identity probabilities  \(\left(\mu (m_{l-1},m_{l})\right)\), and the conditional probabilities \(\left(\mu (m_{l-1}|m_{l})\right)\) become stationary for a long polymer and can be calculated. One must first  calculate the partial velocities, \(v_{r}\) and \(v_{w}\). The quantity \(v_{x}\mu(x)\) is the net rate at which monomers are added after an \(x\),
\begin{align}
 v_{x}=\psi^{+}_{xr}-\frac{\mu(x|r)\mu(r)}{\mu(x)}\psi^{-}_{xr}+\psi^{+}_{xw}-\frac{\mu(x|w)\mu(w)}{\mu(x)}\psi^{-}_{xw}. \label{eq:vx1}
\end{align}
Following ref.~\cite{Gaspard}, these velocities
%
%
can be solved in terms of the propensities. In turn, the velocities and propensities determine
 tip and conditional probabilities $\mu(m_l)$ and $\mu(m_{l-1}|m_l)$ \cite{Gaspard}. Further details are provided in SI Appendix. 
%
%

Gaspard's method describes the chain while it is still growing through the stochastic variables \(M_{l}\) and  \(M_{l-1}\), with the index \(l\) being the current length of the polymer. We, however, are interested in the identity of the monomer at position \(n\) when \(l \gg n\). We label this "final" state of the monomer at position \(n\) as \(M_{n}^{\infty}\). As discussed in the SI Index, \(M_n^\infty\) is distinct from \(M_n\) near the tip. \(M_n^\infty\) is described by the error probability \(\epsilon\) and the conditional error probabilities \(\epsilon_{r}\) and \(\epsilon_{w}\), defined as the probability that \(M^\infty_{n+1} = w\) given that \(M^\infty_{n} = r\) or \(M^\infty_{n} = w\), respectively. 
We show that \(\epsilon\), \(\epsilon_{r}\) and \(\epsilon_{w}\) are sufficient in the SI Appendix, by proving that the \(M^\infty\) is a Markov chain of \(r\) and \(w\) monomers (this requirement is distinct from the Markovian growth dynamics). We note that \(\epsilon\neq\epsilon_{r}\neq\epsilon_{w}\) as a direct result of the dependence of the transition propensities on the current and previous tip monomers, which in turn arises from detachment.

\begin{figure*}
\begin{center}
\includegraphics[scale=0.15]{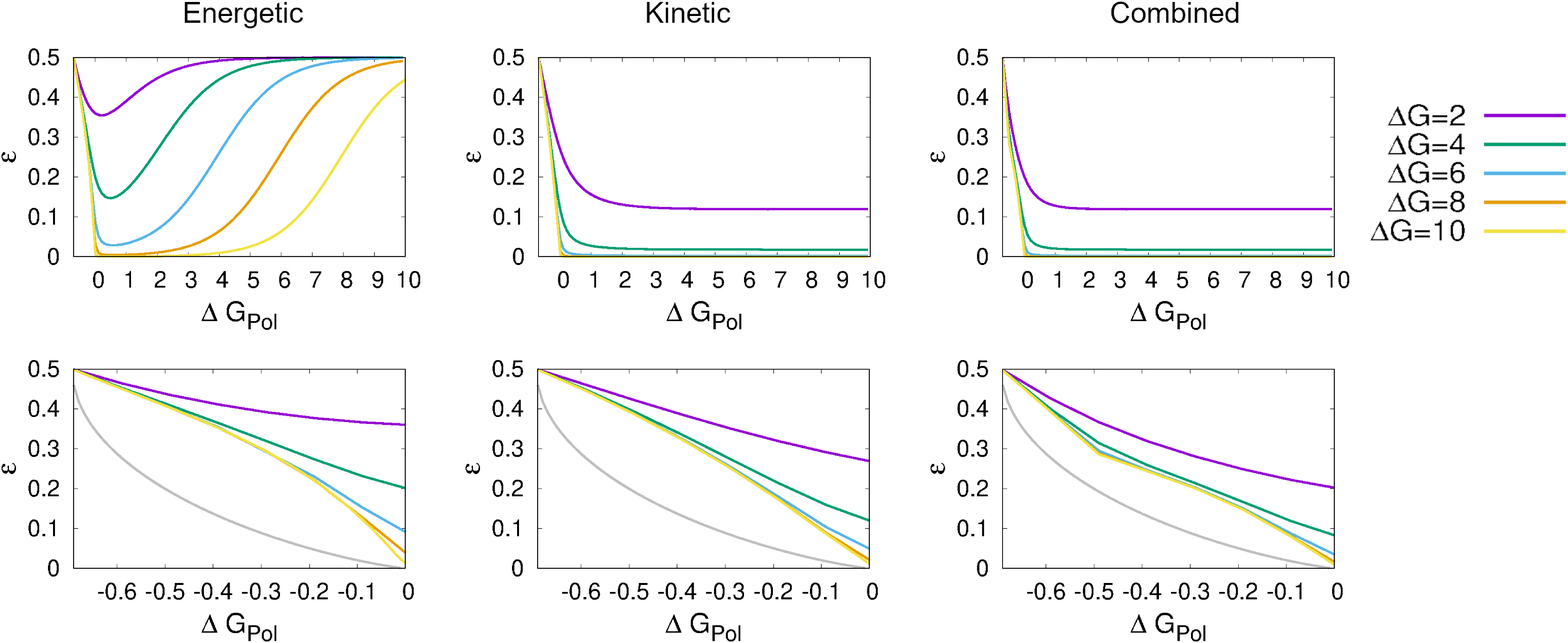}
\caption{Error probability \(\epsilon\) as a function of \(\Delta G_{\rm{\rm{pol}}}\) for all three mechanisms: (a) over a wide range of  \(\Delta G_{\rm{\rm{pol}}}\), and (b) within the entropy-driven region \(\Delta G_{\rm pol} \leq 0\). The temporary thermodynamic mechanism is always least accurate, and the combined mechanism most accurate. All mechanisms have no accuracy in the limit of \(\epsilon \rightarrow 0\), and are far from the fundamental bound on single-site accuracy imposed by \(H_{\rm ss} = -\epsilon \ln \epsilon - (1-\epsilon) \ln (1-\epsilon) \geq -\Delta G_{\rm pol}\) except at \(\Delta G_{\rm pol}\approx 0\).}
\label{Error}
\end{center}
\end{figure*}

To calculate \(\epsilon\), \(\epsilon_r\) and \(\epsilon_w\), we define currents \(J_{xy}\) that are related to both \(\psi^{\pm}_{xy}\) and \(\epsilon\), \(\epsilon_r\) and \(\epsilon_w\) separately. The current \(J_{xy}\) is the net rate per unit time at which transitions \(\&x\rightarrow \&xy\) occur: \(J_{xy}=\psi_{xy}^{+}\mu(x)-\psi_{xy}^{-}\mu(x,y)\). By considering the transitions in our system as a tree, as in Fig. \ref{Flux}, we can relate the current through a branch to the overall rate at which errors are permanently incorporated into a polymer growing at total velocity \(v=v_{r}\mu(r)+v_{w}\mu(w)\)
\begin{align}
\begin{split}
 J_{rr}=(1-\epsilon)(1-\epsilon_{r})v=\mu(r)\psi^{+}_{rr}-\mu(r,r)\psi^{-}_{rr},\label{eq:jrr}
\end{split}\\
\begin{split}
 J_{rw}=(1-\epsilon)\epsilon_{r}v=\mu(r)\psi^{+}_{rw}-\mu(r,w)\psi^{-}_{rw},\label{eq:jrw}
\end{split}\\
\begin{split}
J_{wr}=\epsilon(1-\epsilon_{w})v=\mu(w)\psi^{+}_{wr}-\mu(w,r)\psi^{-}_{wr},\label{eq:jwr}
\end{split}\\
\begin{split}
J_{ww}=\epsilon\epsilon_{w}v=\mu(w)\psi^{+}_{ww}-\mu(w,w)\psi^{-}_{ww}.\label{eq:jww}
 \end{split}
\end{align}

Eliminating \(\epsilon\) from the simultaneous equations \ref{eq:jrr}-\ref{eq:jww} yields \(\epsilon_r\) and \(\epsilon_w\) in terms of known quantities. To find \(\epsilon\), note that the final sequence has a transition matrix parameterized by \(\epsilon_r\) and \(\epsilon_w\), with the overall error \(\epsilon\) given by its dominant eigenvector. As detailed in the SI Appendix, we obtain
%
%
\(    \epsilon = {\epsilon_{r}}/{(1+\epsilon_{r}-\epsilon_{w})}
\).
From \(\epsilon\), \(\epsilon_r\) and \(\epsilon_w\) we calculate properties of the copy in terms of \(\psi^{\pm}_{xy}\) and thus the free energies. We corroborate the analytical results with simulation (see SI Index).

\section*{Results}
\subsection*{General thermodynamic bounds}
The free energy of the combined bath and polymer system decreases over time. There are two contributions to the free-energy change per added monomer: the chemical free energy  \(\Delta G_{\rm{pol}}\), and a contribution from the uncertainty of the final polymer sequence \cite{Ouldridge}. 
The latter is quantified by the entropy rate \(H\) \cite{Crutchfield, Cover}:
\begin{align}
    H({M^\infty})=\lim_{n\rightarrow\infty}\frac{1}{n}H(M^\infty_{1},M^\infty_{2},...,M^\infty_{n}),
\end{align}
which in our case is given by \cite{Cover}
\begin{align}
    H({M^\infty})=&-\epsilon\left(\epsilon_{w}\ln{\epsilon_{w}}+(1-\epsilon_{w})\ln{(1-\epsilon_{w})}\right)\nonumber\\&-(1-\epsilon)\left(\epsilon_{r}\ln{\epsilon_{r}}+(1-\epsilon_{r})\ln{(1-\epsilon_{r})}\right).
\end{align}
The overall free energy change per added monomer is then \(\Delta G_{\rm{tot}}=-\Delta G_{\rm{pol}}-H\), which must be negative for growth: \(H \geq -\Delta G_{\rm pol}\). Since copy-template interactions are not extensive in the copy length, they do not contribute. Given that \(H\geq0\), growth is possible in the region where \(\Delta G_{\rm{pol}}<0\), corresponding to the ``entropically driven'' regime \cite{esposito2010,Bennett}.

The entropy rate is bounded by the single site entropy \(H \leq H_{ss}=-\epsilon\ln{\epsilon}-(1-\epsilon)\ln{(1-\epsilon)}\). \(H_{ss}\) quantifies the desired correlations {\it between} copy and template. For previous models of TSA with uncorrelated monomer incorporation, \(H=H_{ss}\) \cite{Bennett,Cady,Andrieux,Sartori1,Sartori2,Gaspard,esposito2010}.  In our model,  the necessary complexity of \(\psi^{\pm}_{xy}\) generates correlations {\it within} the copy. A stronger constraint on the single site entropy, and hence accuracy, then follows: \(H_{ss}\geq H\geq -\Delta G_{\rm{\rm{pol}}}\).

Fundamentally, a persistent copy is a high free energy state, as the entropic cost of copy-template correlations cannot be counteracted by stabilizing copy-template interactions. Thus the process moves a system between two high free energy states, converting chemical work into correlations. 
In general, only a fraction of the chemical work done by the monomer bath is retained in the final state, implying dissipation, and so it is natural to introduce an efficiency. The overall free energy stored in the polymer has contributions both from the creation of an equilibrium (uncorrelated) polymer and from correlations within the copy and with the template. We are interested only in the contributions above equilibrium. The efficiency \(\eta\) is then the proportion of the additional free energy expended in making a copy above the minimum required to grow a random equilibrium polymer that is successfully converted into the non-equilibrium free energy of the  copy sequence rather than being dissipated. In our simple case, 
\begin{equation}
\eta\equiv\frac{H_{\rm eq}-H}{H_{\rm eq}+\Delta G_{\rm{pol}}} \leq 1. \label{eq:eff}
\end{equation}
Here, \(\Delta G_{\rm{pol}}+H_{\rm eq} = \Delta G_{\rm{pol}}+ \ln2\) is the extra chemical work done by the buffer above that required to grow an equilibrium polymer, \(\Delta G_{\rm pol}^{\rm eq} = -H_{\rm eq} = -\ln 2\). The free energy stored in the copy sequence, above that stored in an equilibrium system, is \(H_{\rm eq}-H\). \(\eta \leq1\) follows from \(\Delta G_{\rm{pol}}+ H \geq 0\). Similarly, since \(H_{\rm ss}\geq H\) we can define a single-site efficiency 
\begin{equation}
\eta_{\rm ss}\equiv\frac{H_{\rm eq}-H_{\rm ss}}{H_{\rm eq}+\Delta G_{\rm{pol}}} \leq \eta \leq 1. \label{eq:effss}
\end{equation}
Unlike \(\eta\), the single site efficiency \(\eta_{\rm ss}\) discounts the free energy stored in  ``useless" correlations {\it within} the copy.

\subsection*{Behaviour of specific systems}
We consider three representative models consistent with eqs.~\ref{eq:rr}-\ref{eq:ww}. First, the purely kinetic mechanism obtained by setting \(\Delta G_{\rm{TT}}=0\) and \(\Delta G_{\rm{K}}=\Delta G\) in  equations \ref{eq:rr}-\ref{eq:ww}. Originally proposed
 by Bennett for TSA~\cite{Bennett}, it is coincidentally a limiting case of persistent copying since there is no equilibrium bias. We also consider two new mechanisms:
pure ``temporary thermodynamic discrimination'' with \(\Delta G_{\rm{K}}=0\) and \(\Delta G_{\rm{TT}}=\Delta G\), and a ``combined discrimination mechanism'', in which both template binding strengths and rates of addition favour \(r\) monomers: \(\Delta G_{\rm{K}}=\Delta G_{\rm{TT}}=\Delta G\).


All three mechanisms have two free parameters, the overall driving strength \(\Delta G_{\rm{pol}}\) and the discrimination parameter \(\Delta G\). We plot error probability against \(\Delta G_{\rm{pol}}\) for various \(\Delta G\) in  fig. \ref{Error}. Also shown is the thermodynamic lower bound on \(\epsilon\) implied by \(H_{\rm ss}\geq H\geq -\Delta G_{\rm{\rm{pol}}}\). All three cases have no accuracy (\(\epsilon=0.5\)) in equilibrium (\(\Delta G_{\rm pol}\rightarrow-\ln2\))
since an accurate persistent copy is necessarily out of equilibrium \cite{Ouldridge}. By contrast, TSA allows for  accuracy in equilibrium \cite{Sartori1,Sartori2,Johansson}.

The temporary thermodynamic mechanism is always the least effective. It has no accuracy for high \(\Delta G_{\rm pol}\) as the difference between \(r\) and \(w\) is only manifest when stepping backwards, and for high \(\Delta G_{\rm pol}\) back steps are rare \cite{Sartori1,Sartori2}. More interestingly, temporary thermodynamic discrimination is also inaccurate as \(\Delta G_{\rm pol}\rightarrow-\ln2\), when the system takes so many back steps that it fully equilibrates. Low \(\epsilon\)  only occurs when \(\Delta G_{\rm{pol}}\) is sufficient to inhibit the detachment of \(r\) monomers, but not the detachment of \(w\) monomers. This trade-off region grows with \(\Delta G\). By contrast, both the combined case and the kinetic case have accuracy in the limit of \(\Delta G_{\rm{pol}}\rightarrow \infty\), since they allow \(r\) to bind faster than \(w\).
 Considering \(\Delta G_{\rm pol}\leq0\) closely (fig. \ref{Error}b) shows the combined case to be superior.

\begin{figure*}
\begin{center}
\includegraphics[scale=0.15]{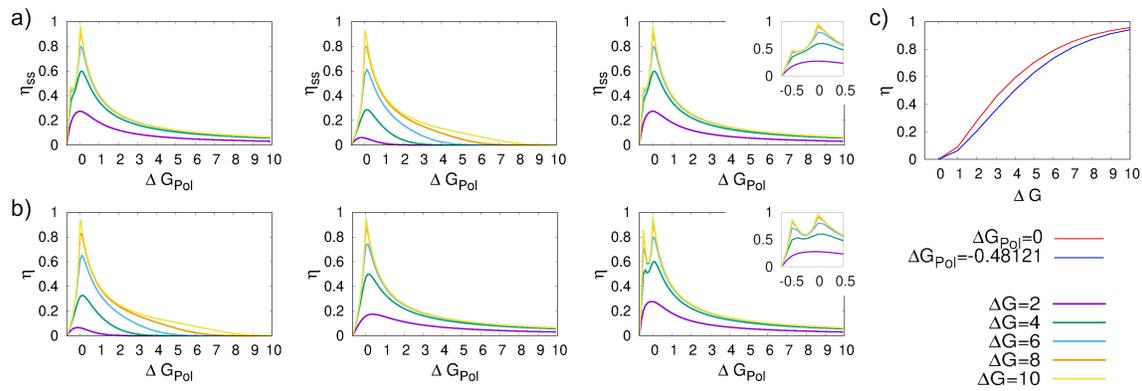}
\caption{Effieciencies \(\eta_{\rm ss}\) (a) and \(\eta\) (b) plotted against \(\Delta G_{\rm pol}\) show sharp peaks at \(\Delta G_{\rm{pol}}=0\) as \(\Delta G\rightarrow \infty\) in all three cases. In the combined case we see a second peak in \(\eta\), and a shoulder in \(\eta_{\rm ss}\) at \(\Delta G_{\rm{pol}}=-0.48121\). (c) Both of these peaks in \(\eta\) tend to unity at \(\Delta G \rightarrow \infty\).}
\label{Efficiency}
\end{center}
 \end{figure*}

Intriguingly, all three mechanisms are far from the fundamental bound on \(\epsilon\) implied by \(H_{\rm ss}\geq H\geq -\Delta G_{{\rm{pol}}}\) as \(\Delta G_{\rm{pol}}\rightarrow-\ln{2}\), and there is an apparent cusp in \(\epsilon\) at \(\Delta G_{\rm pol} \approx 0.48\) as \(\Delta G \rightarrow \infty\) in the combined case.
The performance relative to the bound is quantified by \(\eta_{\rm ss}\) (fig. \ref{Efficiency}). Surprisingly, we observe in fig.~\ref{Efficiency}a that not only does \(\epsilon\) go to zero as \(\Delta G_{\rm pol}\rightarrow-\ln2\), so does \(\eta_{ss}\) in all cases. For small non-equilibrium driving, none of the extra chemical work input is stored in correlations with the template. Mathematically, this inefficiency arises because \(\epsilon-0.5 \propto \Delta G_{\rm{pol}}-\ln{2}\) as \(\Delta G_{\rm pol}\rightarrow-\ln2\) (as observed in fig.~\ref{Error}), and \(H_{\rm ss}-\ln2 \propto (\epsilon-0.5)^{2}\) for \(\epsilon\approx0.5\) by definition. Thus, from equation \ref{eq:effss}, \(\eta_{\rm ss}\propto \Delta G_{\rm{pol}}-\ln{2}\) as \(\Delta G_{\rm pol}\rightarrow-\ln2\). That this result only depends on error probability decreasing proportionally with \(\Delta G_{\rm{pol}}\) for small driving suggests that a vanishing \(\eta_{ss}\) in equilibrium may be quite general.

In all cases, the single-site efficiency \(\eta_{\rm ss}\) increases from 0 at \(\Delta G_{\rm pol}=-\ln 2\) to a peak near \(\Delta G_{\rm pol}=0\), with \(\eta_{\rm ss} \rightarrow 1\) as \(\Delta G\rightarrow\infty\). Beyond this peak, \(\eta_{\rm ss}\) drops as the stored free energy  is bounded by \(\ln{2}\) per monomer.
To understand the peak,  note that for every  \(\Delta G_{\rm{pol}}\leq0\) there is a hypothetical highest accuracy copy with \(\epsilon\) fixed by \(H_{\rm ss}=-\Delta G_{\rm{pol}}\) that is marginally thermodynamically permitted. However, this copy is not usually kinetically accessible. At \(\Delta G_{\rm{pol}}=0\) the marginal copy has 100\% accuracy and, unusually,  all three mechanisms can approach it kinetically, causing a peak. The efficiency  approaches its limit of unity even for moderate values of \(\Delta G\). We note that as \(\Delta G\rightarrow\infty\), growth is slow for \(\Delta G_{\rm pol}\leq 0\): the total number of steps taken diverges.

A related argument explains the apparent cusp in the error \(\epsilon\) for the combined mechanism at \(\Delta G_{\rm{pol}}\approx-0.48\) and high \(\Delta G\). On the plot of \(\eta_{\rm ss}\) (fig.~\ref{Efficiency}\,a) this cusp manifests as a shoulder. The full efficiency \(\eta\) (fig.~\ref{Efficiency}\,b) has a prominent second peak. Uniquely, the combined mechanism's kinetics strongly disfavour chains of consecutive \(w\)s for high \(\Delta G\). A final copy with no consecutive \(w\)s  has \(\epsilon_{w}=0\) but \(\epsilon_{r}\neq0\). Maximizing the entropy rate of such a Markov chain over \(\epsilon_r\) gives \(H_{\rm max}= 0.48121\); \(\Delta G_{\rm pol}= -H_{\rm max}\) matches the location of our peak/cusp. Thus the combined mechanism initially eliminates consecutive \(w\)s, and at \(\Delta G_{\rm pol} =-0.48121 \) a state with \(\eta_w=0\) is thermodynamically permitted for the first time. For large \(\Delta G\), this polymer is kinetically accessible and grows with thermodynamic efficiency approaching unity (fig.~\ref{Efficiency}\,c). In this limit, the overall entropy generation is zero.


The above behaviour is a striking example of correlations being generated within the copy sequence, as well as with the template. Notably, whilst \(\eta\) approaches unity at this point, \(\eta_{\rm ss}\) does not (fig.~\ref{Efficiency}). Correlations {\it within} the copy sequence limit the chemical work that can be devoted to improving the single site accuracy of the copy polymer, since they prevent the bound \(H_{\rm ss}\geq H\geq -\Delta G_{{\rm{pol}}}\) from being saturated. 

\section*{Conclusion}
The thermodynamic constraint on copying that underlies this work is deceptively simple: unlike TSA, the overall chemical contribution to the free-energy of a copy must be independent of the match between template and copy sequences. By studying the simplest mechanisms satisfying this constraint, however, we can draw important conclusions for copying mechanisms and thermodynamics more generally.

For copying, the most immediate contrast with  previous work on TSA \cite{Bennett,Cady,Andrieux,Sartori1,Sartori2,Gaspard,esposito2010} is that accuracy is necessarily zero when the copy assembles in equilibrium, since equilibrium correlations between physically separated polymers are impossible~\cite{Ouldridge}. Consequently, unlike self-assembly, no autonomous copying system can rely on relaxation to  near-equilibrium; fundamentally different paradigms are required.   

A direct result of the temporary nature of thermodynamic discrimination in persistent copying is that relying solely on the strong binding of correct copies is ineffective in ensuring accuracy. At \(\Delta G= 6k_BT\),  comparable to the cost of a mismatched base pair \cite{SantaLucia2004}, the temporary thermodynamic discrimination model never improves upon \(\epsilon = 0.0285\), which is more than ten times the equilibrium error probability based on energetic discrimination obtainable in TSA, \(1/(1+\exp(\Delta G))\). This performance would degrade further if many competing monomers were present. Mechanisms for copying must therefore be more carefully optimized than those for TSA. Either some degree of direct ``kinetic'' discrimination (with correct monomers incorporated more quickly), or as an alternative, fuel-consuming proofreading cycles, appear necessary. It is well-established that proofreading cycles can enhance discrimination above equilibrium in TSA  \cite{Sartori2,Bennett,EHRENBERG1980333}, and the challenges in achieving direct kinetic discrimination in diffusion-influenced reactions via the details of the microscopic sub-steps may explain the ubiquity of such cycles in true copying systems.



Correlations within the copy, as well as between copy and template, arise naturally in persistent mechanisms. Indeed, in one case, pairs of mistakes are eliminated well before individual mistakes. These correlations contribute to the non-equilibrium free energy of the final state, reducing the single-site copy accuracy achievable for a given chemical work input. Biologically, however, it is arguably the accuracy of whole sequences, rather than individual monomers, that matters. In this case, positive correlations could advantageously  increase the number of 100\% correct copies for a given average error rate. It remains to be seen whether this tactic, which indeed may arise in real systems \cite{Rao}, is feasible. Regardless, we predict that within-copy correlations may be significant, particularly in simple systems with low accuracy. These correlations may change significantly if the requirement to remain bound with one exactly bond is relaxed, and exploiting correlations to extend functionality  beyond simple copying is an intriguing prospect. 

Relaxing this requirement also raises the possibility of early copy detachment. This risk is likely to be worse for genome replication than for transcription and translation, which may explain why the latter proceed by mechanisms analogous to our model, while DNA replication does not: here, the copy of a single DNA strand from the double helix is first completely assembled on the template and separated only much later.

Thermodynamically, a persistent copying mechanism  converts the high free energy of the input molecules into a high free-energy copy state; we have defined a general efficiency of this free-energy transduction for copying. In typical physical systems with tight coupling of reactions, high efficiency occurs when the load is closely matched to the driving, either in autonomous systems operating near the stall force, or in quasistatically manipulated systems. For the polymer copying mechanisms studied here, however, we find that the thermodynamic efficiency of information transfer, and not just the accuracy, approaches zero as polymer growth stalls. We predict that this result is general, since the alternative would require a sub-linear convergence of the error rate on 50\% as thermodynamic driving tends towards the stall point. 

Fundamentally, the copy process transduces free energy into a complex system with many  degrees of freedom (the sequence), and not just the polymer length. To be accurate, the sequence must be prevented from equilibrating. Thus, whilst weak driving leads to polymer growth with little overall entropy generation, it does a poor job of pushing the polymer sequence out of equilibrium. We predict that similar behaviour will arise in other systems intended to create an output in which a subset of the degrees of freedom are out of equilibrium.

Away from the equilibrium limit, the efficiency shows one or more peaks as the polymerization free energy is varied. At these peaks, the system transitions between two non-equilibrium states with remarkably little dissipation. These particular values of \(\Delta G_{\rm pol}\) are sufficient to stabilise non-equilibrium distributions that happen to be especially kinetically accessible, rendering the true equilibrium particularly inaccessible. This alignment of kinetic and thermodynamic factors is most evident in the combined mechanism that efficiently produces a state with few adjacent mismatches  at \(\Delta G_{\rm pol} = -0.48121\). These results slightly qualify the prediction of Ref. \cite{Ouldridge} that accurate copying is necessarily entropy generating, since entropy generation can be made arbitrarily small, whilst retaining finite copy accuracy, by taking \(\Delta G\rightarrow \infty\) at these specific values of \(\Delta G_{\rm pol}\).

The behaviour of the efficiency in these models emphasizes the importance, in both natural and synthetic copying systems, of kinetically preventing equilibration. Our work emphasizes that this paradigm should be applied not only to highly-evolved systems with kinetic proofreading mechanisms \cite{Hopfield}, but also the most basic mechanisms imaginable. 

Extending our analysis to consider fuel-consuming proofreading cycles would be natural. However these cycles will not change the fundamental result that the entropy of the copy sequence is thermodynamically constrained, in this case by \(H_{\rm ss} \geq H \geq -\Delta G_{\rm pol} - \Delta G_{\rm fuel}\), where the final term is the additional free energy expended per step to drive the system around proofreading cycles. We predict that non-equilibrium proofreading cycles, by their very nature, are unlikely to approach efficiencies of unity.

We thank Charles Bennett for instructive conversation. TEO is supported by the Royal Society and PRtW is supported by the Netherlands Organisation for Scientific Research.

\section*{Supplementary Information}

\subsection*{Solving for partial velocities and tip probabilities}
Following ref.~\cite{Gaspard}, the partial velocities
can also be expressed as
\begin{align}
 v_{x}=\frac{\psi^{+}_{xr} v_{r}}{\psi^{-}_{xr}+v_{r}} +\frac{\psi^{+}_{xw} v_{w}}{\psi^{-}_{xw}+v_{w}}, \label{eq:vx}
\end{align}
for \(x=r,w\). This gives a pair of simultaneous equations that 
can be solved for the partial velocities in terms of the propensities. In turn, the velocities and propensities determine tip and conditional probabilities $\mu(m_l)$ and $\mu(m_{l-1}|m_l)$, with $m_l, m_{l-1} \in \{r,w\}$, via
\begin{align}
\begin{split}
 \mu(x)=\frac{\psi^{+}_{rx}}{\psi^{-}_{rx}+v_{x}}\mu(r) +\frac{\psi^{+}_{wx} }{\psi^{-}_{wx}+v_{x}}\mu(w), \label{eq:mux}
\end{split}\\
\begin{split}
 \mu(x|y)=\frac{\psi^{+}_{xy}}{\psi^{-}_{xy}+v_{y}}\frac{\mu(x)}{\mu(y)}. \label{eq:muxy}
\end{split}
\end{align}

 \subsection*{Difference between the tip probabilities and the final sequence}
 It might not be immediately obvious why the properties of the growing chain described by \(\mu(m_l)\) and \(\mu(m_l,m_{l-1})\) should be different from those of the final chain described by \(\epsilon\), \(\epsilon_{r}\) and \(\epsilon_{w}\), but the difference can be illustrated with a simple example. Consider a system in which incorrect monomers could be added to the end of the chain, but where nothing can be added after an incorrect match. In this case while the tip probability for an incorrect match \(\mu(w)\) would be finite, the error of the final chain \(\epsilon\) would be vanishingly small, as all incorrect matches would have to be removed in order for the polymer to grow further.
 
\subsection*{Demonstrating that the sequence of the final chain is Markovian}
 
Let \(M^\infty_{n}\) be the monomer at the \(n\)th site in the final chain. Let a polymer be represented by \(M^\infty_{1},...,M^\infty_{n}\). The probability of a given chain existing is then \(P^\infty(m_{1},...,m_{n})\). In order for the sequence of monomers moving along the chain (increasing \(n\)) to be able to be represented by a Markov chain, the condition 
\begin{equation}
 P^\infty(m_{n}|m_{n-1},...,m_{1})=P^\infty(m_{n}|m_{n-1})
\label{eq:M1}
\end{equation}
must hold.

In order to demonstrate that eq.~\ref{eq:M1} holds, we rewrite the final chain probability in terms of properties of the growing chain. Specifically we state that the probability of the sequence \(m_{1},...,m_{n}\) existing in the final chain is the product of the probability \(Q(m_{1},...,m_{n-1},t)\) that the chain is in the state \(m_{1},...,m_{n-1}\) at a time \(t\) during the growth process, and the propensity \(\nu(m_n;m_{1},...,m_{n-1})\) with which \(m_{n}\) is added to a chain \(m_{1},...,m_{n-1}\) and never removed, integrated over all time. It should be noted that \(r(m_n;m_{1},...,m_{n-1})\) is time-independent.

\begin{equation}
P^\infty(m_{1},...,m_{n})=\int Q(m_{1},...,m_{n-1},t)\nu(m_{n};m_{1},...,m_{n-1})dt
\label{eq:M2}
\end{equation}
\begin{equation}
P^\infty(m_{1},...,m_{n})=\nu(m_{n};m_{1},...,m_{n-1}) \int Q(m_{1},...,m_{n-1},t)dt
\label{eq:M3}
\end{equation}

Setting the integral equal to \(I(m_{1},...,m_{n-1})\) gives

\begin{equation}
P^\infty(m_{1},...,m_{n})=\nu(m_{n};m_{1},...,m_{n-1}) I(m_{1},...,m_{n-1})
\label{eq:M4}
\end{equation}

Let's consider the probabilities of two sub-sequences, identical except for the final monomer. We call the two final monomers \(m_{n}\) and \(m^\prime_{n}\) and we can denote the ratios of the probabilities of these two chains as follows.
\begin{equation}
\frac{P^\infty(m_{1},...,m_{n})}{P^\infty(m_{1},...,m^\prime_{n})}=\frac{\nu(m_{n};m_{1},...,m_{n-1})I(m_{1},...,m_{n-1})}{\nu(m^\prime_{n};m_{1},...,m_{n-1})I(m_{1},...,m_{n-1})}
\label{eq:M5}
\end{equation}
The \(I\) terms are independent of this final monomer and so cancel. Thus
\begin{equation}
\frac{P^\infty(m_{1},...,m_{n})}{P^\infty(m_{1},...,m^\prime_{n})}=\frac{\nu(m_{n};m_{1},...,m_{n-1})}{\nu(m^\prime_{n};m_{1},...,m_{n-1})}
\label{eq:M6}
\end{equation}

The same relationship holds for the conditional probabilities
\begin{equation}
\frac{P^\infty(m_n|m_{1},...,m_{n-1})}{P^\infty(m^\prime_{n}|m_{1},...,m_{n-1})}=\frac{r(m_{n};m_{1},...,m_{n-1})}{r(m^\prime_{n};m_{1},...,m_{n-1})}
\label{eq:M7}
\end{equation}

For our system, the propensity with which a monomer \(m_{n}\) is added and never removed, \(\nu(m_{n};m_{1},...,m_{n-1})\), is dependent on only on the final two monomers in the sequence fragment, \(m_n\) and \(m_{n-1}\). To see why, note that this propensity is determined by addition and removal of monomers at sites \(i \geq n\). The identities of monomers at positions \(j<n-1\), however, only influence addition and removal propensities at sites \(k<n\) (eq.~1-4 main text).  
Thus we convert \(\nu(m_{n};m_{1},...,m_{n-1})\) to \(f(m_{n},m_{n-1})\).

\begin{equation}
\frac{P^\infty(m_n|m_{1},...,m_{n-1})}{P^\infty(m^\prime_{n}|m_{1},...,m_{n-1})}
=\frac{f(m_{n},m_{n-1})}{f(m^\prime_{n},m_{n-1})}
\label{eq:M8}
\end{equation}

Multiplying both sides by \(P^\infty(m_{1},...,m_{n-2})\) and summing over all values of \(m_{1},...m_{n-2}\) (recall \(\sum_{c,d}P(a|b,c,d)P(c,d)=P(a|b)\) and \(\sum_{c,d}P(c,d)=1\)) gives
\begin{equation}
\frac{P(m_{n}|m_{n-1})}{P(m^\prime_{n}|m_{n-1})}=\frac{f(m_{n},m_{n-1})}{f(m^\prime_{n},m_{n-1})}.
\label{eq:M9}
\end{equation}
Comparing equations \ref{eq:M8} and \ref{eq:M9} yields:
\begin{equation}
\frac{P^\infty(m_{n}|m_{n-1},...,m_{1})}{P(m^\prime_{n}|m_{n-1},...,m_{1})}=\frac{P^\infty(m_{n}|m_{n-1})}{P^\infty(m^\prime_{n}|m_{n-1}).}
\label{eq:M10}
\end{equation}

Summing over the possible values of \(m^\prime_{n}\) and recalling that \(P^\infty(r|m_{n-1})+P^\infty(w|m_{n-1})=1\) and \(P^\infty(r|m_{n-1},...,m_{1})+P^\infty(w|m_{n-1},...,m_{1})=1\) yields:
\begin{equation}
\begin{split}
P^\infty(m_{n}|m_{n-1},...,m_{1})= P^\infty(m_{n}|m_{n-1}),
\label{eq:M11}
\end{split}
\end{equation}
thereby proving that the sequence of the final chain is Markovian.

\subsection*{Overall error probability of the final chain}

\begin{figure}
\begin{center}
\includegraphics[scale=0.3]{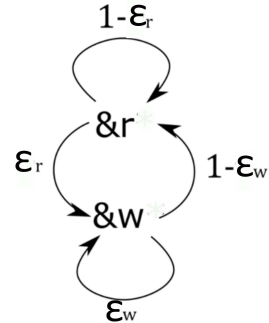}
\end{center}
\caption{The transition diagram for the Markov process describing the sequence of monomers found by stepping forward along a completed chain.}
\label{Trans}
\end{figure}

The final sequence is described by the Markov chain illustrated in figure \ref{Trans}. The transition matrix for this process is
\begin{align}
T=\begin{bmatrix}
   1-\epsilon_{r}       &1-\epsilon_{w} \\
    \epsilon_{r}       & \epsilon_{w} \\ 
\end{bmatrix},\label{eq:matrix}
\end{align}
The eigenvector of this transition matrix with eigenvalue equal to unity gives the steady state of the Markov chain. The second component of this eigenvector corresponds to the overall probability of incorrect matches, \(\epsilon\):
\begin{align}
    \epsilon = {\epsilon_{r}}/{(1+\epsilon_{r}-\epsilon_{w})}.
\end{align}

\begin{figure*}
\includegraphics[scale=0.4]{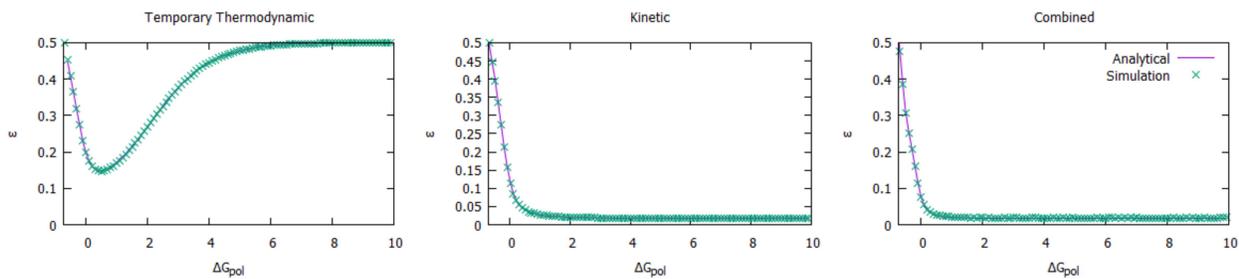}
\caption{Errors obtained from Gillespie simulations are indistinguishable from the analytical results obtained using Gaspard's method for all three mechanisms.}
\label{compareerr}
 \end{figure*}
 

\subsection*{Corroboration with simulation}
To check the analytical methods used to solve the system we also simulated the growth of a polymer. We used a Gillespie simulation \cite{Gillespie}, with transition rates given by \(\psi^\pm_{xy}\). Simulations were initialised with a randomly determined two monomer sequence, and truncated as soon as the polymer reached 1000 monomers. We found that such a length rendered edge effects negligible in all but the most extreme cases for the calculation of \(\epsilon\). Polymer error probabilities were inferred directly from the 100 simulations, and are compared to analytical results in fig. \ref{compareerr}.

We note in passing that the calculation of \(H\), \(H_{ss}\), and particularly the efficiency's \(\eta\) and \(\eta_{ss}\), are more vulnerable to random fluctuations in a simulation of finite length, and peculiar edge effects, than \(\epsilon\). Gaspard's solution is therefore invaluable in reaching robust conclusions for these quantities.

\end{document}